\documentclass[dvips]{article}
\usepackage{icrctc07}

\title{Observation of GRBs by the MAGIC Telescope, Status and Outlook}
\shorttitle{Dependence of the energy}
\authors{D. Bastieri$^{1}$, N. Galante$^{2}$, M. Garczarczyk$^{2}$, M. Gaug$^{3}$, F. Longo$^{4}$, S. Mizobuchi$^{2}$ and \mbox{V. Scapin}$^{5}$ for the MAGIC Collaboration}
\shortauthors{G. A. Medina-Tanco and et al}
\afiliations{$^1$  Dipartimento di Fisica - Universit\`a di Padova \& INFN Padova, Via F. Marzolo 8, I-35131 Padova, Italy\\ $^2$ Max-Planck-Institut f\"ur Physik, F\"ohringer Ring 6, D-80805 Munich, Germany\\  $^3$ Instituto de Astrof\'isica de Canarias, Calle V\'ia L\'actea, E-38205 La Laguna, Tenerife, Spain\\$^4$ Dipartimento di Fisica - Universit\`a di Trieste \& INFN Trieste, Via F. Valerio 2, I-34127 Trieste, Italy\\ $^5$ Dipartimento di Fisica - Universit\`a di Udine \& INFN Trieste, Via delle Scienze 208, I-33100 Udine, Italy}
\email{garcz@mppmu.mpg.de}

\abstract{Observation of Gamma Ray Bursts (GRBs) in the Very High Energy (VHE) domain will provide important information on the physical conditions in GRB outflows. The MAGIC telescope is the best suited Imaging Atmospheric Cherenkov Telescope (IACT) for these observations. Thanks to its fast repositioning time and low energy threshold, MAGIC is able to start quickly the follow-up observation, triggered by an alert from the GRB Coordinates Network (GCN), and observe the prompt emission and early afterglow phase from GRBs. In the last two years of operation several GRB follow-up observations were performed by MAGIC, however, until now without successful detection of VHE $\gamma$-rays above threshold energies $>100 \, \mathrm{GeV}$. In this paper we revise the expectations for the GRB observations with MAGIC, based on the experience from the last years of operation.}

\begin{document}
\maketitle

\section{Introduction}

There are several competing models describing the prompt and the afterglow emission phase of GRBs. In order to unravel the true nature of the prompt emission as well as to constrain important physical quantities such as the bulk Lorentz factor and the magnetic fields in the outflow, more broadband observations including the $\mathrm{GeV}-\mathrm{TeV}$ bands are needed.

Due to the difficulties of the observation of these short lasting phenomena, the spectra of GRBs could not yet be measured above $\sim 20 \, \mathrm{GeV}$, the maximal energy of the size limited space born detectors. However, there are hints for a VHE component from GRBs: The EGRET detector observed delayed GeV emission from GRB940217~\cite{hurley}. Also in the case of GRB941017 a separate high energy component with a spectral index of $\beta = -1$ was observed~\cite{gonzales}. And finally a tentative detection of radiation above $100 \, \mathrm{GeV}$ from GRB970417a was made by \mbox{MILAGRITO}~\cite{atkins}. Apart from these measurements only upper limits on the VHE emission could be obtained.

With a detector like MAGIC~\cite{magic} the situation has improved. MAGIC is currently the largest IACT in the world and was built to explore the $\gamma$-ray sky with high sensitivity at energies starting well below $100 \, \mathrm{GeV}$. Moreover, the lightweight design of its supporting cradle allows MAGIC to slew to any position in the sky in less than $100 \, \mathrm{s}$ (on average within $45 \, \mathrm{s}$). These specifications make MAGIC the best ground based experiment able to observe the GRB prompt emission and early afterglow phase from GRBs.

Having a small field of view, IACTs rely on an external trigger, as the one provided by the automated satellite link to the GCN\footnote{$\mathrm{http://gcn.gsfc.nasa.gov/}$}. It broadcasts the coordinates of events triggered and selected by dedicated satellite detectors and sends them to ground based experiments. The {\it Swift} BAT detector is currently the most active GRB triggering facility and provides the majority of the alerts for the MAGIC telescope. This paper concentrates on the technical aspects of MAGIC observations, results are reported in a separate contribution~\cite{magic.results}. 

\section{Sensitivity of the MAGIC telescope for GRB observations}

The performance of the MAGIC telescope and the sensitivity of the data analysis were tested with the data set taken from the Crab Nebula, the standard candle in the VHE range. The typical GRB observational conditions (interruption of the ongoing observation, fast repositioning towards the new coordinates, refocussing of the mirror elements and calibration of the telescope camera) were guaranteed by an accidental alert from the INTEGRAL satellite on October $11^{\mathrm{th}}$, 2005. Until the accidental trigger from the Crab Nebula was realized~\cite{mereghatti}, $2814 \, \mathrm{s}$ of data were recorded by MAGIC already. The standard approach used for the analysis of GRB data yielded a $14 \sigma$ signal above $350 \, \mathrm{GeV}$~\cite{scapin}. Based on this result we extracted the sensitivity of our analysis: MAGIC can detect  a source of 5 Crab Unit (C.U.)\footnote{$\mathrm{C.U.} = 1.5 \cdot 10^{-3} \left( \frac{E}{\mathrm{GeV}} \right)^{-2.59} \frac{\mathrm{ph}}{\mathrm{cm^{2} \, s \, TeV}}$} intensity at $5 \sigma$ level within $40 \, \mathrm{s}$, in the energy band above $300 \, \mathrm{GeV}$, and within $90 \, \mathrm{s}$ at energies below $300 \, \mathrm{GeV}$.

The energy threshold of an IACT scales with the observation zenith angle $Zd$. A threshold energy of $80 \, \mathrm{GeV}$ at Zenith increases to $520 \, \mathrm{GeV}$ at $Zd=60^{\circ}$. This fact is relevant since most of the GRB events are situated at redshift $z>1$, for which photons above $100 \, \mathrm{GeV}$ get absorbed by the Extragalactic Background Light (EBL)~\cite{kneiske}. As the redshift of the source is normally only known after several days from optical follow-up observations, one is obliged to observe all candidates, even though some measurements may subsequently turn out useless in the light of the object's redshift.

\section{Summary of the GCN activity}

The MAGIC facility uses an internet socket connection to communicate with the GCN. A demon program performs a full-time survey of the alerts and validates them with the predefined observability criteria:
\begin{itemize}
\item sun below the astronomical horizon
\item angular distance to the Moon $> 30^{\circ}$
\item zenith angle $< 60^{\circ}$
\end{itemize}

\noindent
If the above mentioned criteria are fulfilled and the weather conditions allow the observation, the GRB coordinates are sent to the central control software. In the next step the shift crew needs to accept the new observation, which initiates the fast repositioning towards the GRB sky coordinates and starts data taking.

Between January 2005 and May 2007 in total 352 acceptable alerts were received from GCN, out of which 276 came from {\it Swift} BAT. From the latter one, 33 events were later declared as not due to a GRB, which results in a fake alert rate of $12 \%$ and a GRB frequency of $\sim9.7$ GRBs / month, before application of the observability criteria.

\section{MAGIC GRB observations}

The MAGIC telescope responded to 23 of the {\it Swift} BAT alerts and started the follow-up observation, corresponding to a duty cycle of $8.3 \%$ and an observation frequency of $\sim 1$ GRB / month. Table~\ref{tab:magicgrbs} summarizes these observations and gives technical details of the observation parameters. 

It has to be mentioned that for the first five bursts the fast movement mode of the telescope was not yet implemented, resulting in longer repositioning times and delays to the burst onset $T_{0}$. Furthermore three of the MAGIC observations were triggered by a fake alert and are not included in the table.

In order to reduce further the response time of the telescope, the requested explicit confirmation by the shift crew to enter the fast movement mode, is going to be eliminated soon. Additional improvement will be achieved by allowing the turn of the telescope over Zenith. In this way long azimuth distances, which are currently most limiting the repositioning time, will be reduced. 

\begin{table*}[th!]
\begin{center}
\begin{tabular}{r|c|c|c|c|c|c|c|c}
 & GRB & $z$ & $T_{90} \, [\mathrm{s}]$ & $\Delta \mathrm{Az}$ $[\mathrm{deg}]$ & rep. time $[\mathrm{s}]$ & Zd [$\mathrm{deg}$] & start obs. $[\mathrm{s}]$ & obs. time $[\mathrm{min}]$ \\ \hline
 1. & 050421 & -- & 10.3  & 30 & 26 & 52 & 108 & 76\\
 2. & 050505 & 4.27 & 60.0 & 114 & 90 & 49 &  717 & 101\\
 3. & 050509a & -- & 11.6  & 128 & 108 & 58 &131& 119\\
 4. & 050509b & 0.23 & 0.04 & 86 & 83 & 70 &108 & 8\\
 5. & 050528 & -- & 10.8 & 7 & 12 & 49 &77 & 28\\
 6. & 050713a & -- & 70 & 50 & 17 & 49 & 40 & 37 \\
 7. & 050904 & 6.29 & 225 & 55 & 54 & 24 &145 & 147\\
 8. & 060203 & -- & 60 & 268 & 84 & 44 & 268 & 43\\
 9. & 060206 & 4.05 & 7 & 44 & 35 & 13 &59 & 49\\
 10. & 060522 & 5.11 & 69 & 64 & 58 & 60 & 2786 & 13\\
 11. & 060602b & -- & 9 & 102 &191 & 60 & 4109 & 26\\
 12. & 060825 & --  & 8.1 & 71 & 30 & 28 & 57 & 33\\
 13. & 060904a & --  & 80 & 14 & 26 & 60 & 5434 & 119\\
 14. & 060912 & 0.94  & 5.0 & 127 & 40 & 60 & 24291 & 18\\
 15. & 060926 & 3.21& 8.0 & 240 & 83 & 31 & 12923 & 23\\
 16. & 061028   & -- & 106 & 51 & 22 & 30 & 169 & 100\\
 17. & 061110b & 3.44 & 128 & 196 & 69 & 43 & 698 & 59\\
 18. & 061217 & 0.83 & 0.30 & 126 & 45 & 60 & 786 & 66\\
 19. & 070411 & 2.95 & 101 & -- & -- & 38 & 2652 & 128\\
 20. & 070412 & -- & 34 & 26 & 9 & 23 & 701 & 124\\
\end{tabular}
\caption{Summary of all {\it Swift} triggered GRB observations performed by MAGIC since beginning of 2005. The columns from left: burst number, redshift $z$ (if measured), burst duration $T_{90}$, repositioning distance of MAGIC in azimuth direction, required repositioning time, zenith distance at the beginning of the observation, begin of the MAGIC observation in respect to the burst onset $T_{0}$ and the total observation time with MAGIC.}\label{tab:magicgrbs}
\end{center}
\end{table*}

\section{GRB characteristics in the "{\it Swift} era"}

The characteristics of the GRB triggering BAT detector, onboard of the {\it Swift} satellite, has several consequences:

{\bf X-Ray Flashes:} One third of the total GRB population are soft GRBs, also called X-Ray Flashes (XRF)~\cite{heise}. XRFs show similar spectral behavior as GRBs, characterized by the so called Band function~\cite{band}. The only difference is a lower peak energy in the case of the XRF. The narrow energy band of BAT ($15 - 150 \, \mathrm{keV}$) implies that the classification between these two classes cannot be done directly: The peak energy of the XRFs is around $E_{\mathrm{p}} = 36 \, \mathrm{keV}$~\cite{alessio}, while the mean peak energy for GRBs is $E_{\mathrm{p}} = 300 \, \mathrm{keV}$~\cite{kaneko}. There are several scenarios which describe the properties of XRF. However, all these models do not predict VHE $\gamma$-ray emission.

We used the approach presented in~\cite{gendre} to pinpoint possible XRF candidates observed by MAGIC. The method separates XRFs from GRBs, based on the spectral index $\beta$ of the prompt emission phase measured by BAT. With the optimistic cut of $\beta > 2$ we found 37 XRF candidates among the GRBs measured by {\it Swift} up to May $18^{\mathrm{th}}$, 2007 (225 events in total\footnote{$\mathrm{http://swift.gsfc.nasa.gov/docs/swift/archive/grb\_table/}$}). The comparison of the XRFs with the events observed by MAGIC yielded the following candidates as probable XRFs: GRB050509a, GRB050528 and GRB060926. This number is compatible with the statistical expectation.

\vspace{5mm}

{\bf GRB redshift distribution:} The {\it Swift} BAT detector triggers a fainter burst population than was previously the case. The mean redshift of pre-{\it Swift} bursts lay at $z_{\mathrm{mean}} = 1.4$, while bursts discovered by {\it Swift} now have $z_{\mathrm{mean}} = 2.1$. These bursts are extremely useful probes to explore the high-redshift Universe. However, their observability with ground based experiments is strongly limited due to absorption of VHE $\gamma$-rays through pair production with photons of the EBL, when traveling long distances. Out of the 72 {\it Swift} GRBs with measured redshift, only 27 candidates are located at $z < 1$, which is the theoretically maximal distance from which VHE $\gamma$-rays of $100 \, \mathrm{GeV}$ can be observed~\cite{kneiske}. The MAGIC sample contains 7 bursts with redshifts beyond the limiting value. The spectrum of these bursts is expected to show a cutoff at energies around $10 \, \mathrm{GeV}$, excluding a positive detection by MAGIC. The bursts GRB050509b and GRB061217 have redshifts below one, their observation, however, was performed at high zenith angles, resulting in a threshold energy ($E_{\mathrm{th}} > 350 \, \mathrm{GeV}$), again above the cutoff due to EBL absorption.

\vspace{5mm}

{\bf X-ray flares during early afterglow:} Parallel to the {\it Swift} discovery of large redshift GRBs the observations revealed new insights into the early afterglow phase. The discovery of X-ray flares suggests that the GRB engines are active well beyond the observed prompt emission, in some cases on time-scales of up to $\sim 10^{5} \, \mathrm{s}$ or a day~\cite{lazzati}. The study of the temporal properties of flares allows to understand the shock mechanisms. The chance of simultaneous emission of VHE photons together with the X-rays increases. Covering the X-ray flare with MAGIC observations can help to distinguish between the emission processes.

\section{Conclusions}

MAGIC was able to observe part of the prompt and the early afterglow emission phase of many GRBs as a response to the alert system provided by the GCN. The strongest limitation for a detection of VHE $\gamma$-rays are the cosmological distances of the GRBs. In the last two years two bursts with $z < 1$ were observed by MAGIC, although not under optimal observational conditions. We have shown that MAGIC can easily detect sources at $5 \, \mathrm{C.U.}$ level in $90 \, \mathrm{s}$ in its GRB observation mode. Compared with other Cherenkov facilities, MAGIC has the lowest energy threshold and fastest reaction time. For these reasons, MAGIC is currently the best GLAST partner at VHE for prompt GRB observations. In the next future, with the construction of a second telescope, enhancing the sensitivity of MAGIC, the situation is going to improve further, but what is really needed for observation at VHE are close-by GRBs.

\section{Acknowledgements}
The construction of the MAGIC telescope was mainly made possible by the support of the German BMBF and MPG, the Italian INFN, and the Spanish CICYT, to whom goes our grateful acknowledgment. We would also like to thank the IAC for the excellent working conditions at the Observatorio del Roque de los Muchachos in La Palma. This work was further supported by ETH Research Grant TH 34/04 3 and the Polish MNiI Grant 1P03D01028.


\end{document}